\documentclass[runningheads]{llncs}

\usepackage{subfigure}
\usepackage{color}
\usepackage{xcolor}
\usepackage{booktabs}
\usepackage{graphicx}
\usepackage{amsmath,amssymb,amsfonts}
\usepackage{algorithmic}
\usepackage{graphicx}
\usepackage{xspace}
\usepackage{bbold}
\usepackage{textcomp}
\usepackage{mathtools}
\usepackage{bbm}
\usepackage{adjustbox}
\usepackage{arydshln}
\usepackage{pifont}
\usepackage{tikz}
\usepackage{pgfplots}
\pgfplotsset{compat=1.18}
\usepackage{multirow}
\usepackage[normalem]{ulem}
\usepackage[colorlinks=true,linkcolor=blue,citecolor=blue,urlcolor=blue]{hyperref}
\usepackage{breakurl}
\usepackage{marvosym}

\begin{document}
%


\title{Fundus-Enhanced Disease-Aware Distillation Model for Retinal Disease Classification from OCT Images}

\titlerunning{FDDM}
%
\author{Lehan Wang\inst{1} \and
Weihang Dai\inst{1}\and
Mei Jin\inst{2} \and
Chubin Ou\inst{3} \and
Xiaomeng Li\inst{1}$^($\textsuperscript{\Letter}$^)$
}

\authorrunning{L. Wang et al.}


\institute{The Hong Kong University of Science and Technology, Hong Kong, China \\
\email{eexmli@ust.hk}\\
\and
Department of Ophthalmology, Guangdong Provincial Hospital of Integrated Traditional Chinese and Western Medicine, Guangdong, China \\
\and
Guangdong Weiren Meditech Co., Ltd
}


\maketitle              

\def\eg{\emph{e.g.}}
\def\ie{\emph{i.e.}}
\def\etal{\emph{et al. }}
\def\vs{\textit{v.s. }}

\begin{abstract}
Optical Coherence Tomography (OCT) is a novel and effective screening tool for ophthalmic examination. 
Since collecting OCT images is relatively more expensive than fundus photographs, existing methods use multi-modal learning to complement limited OCT data with additional context from fundus images. 
However, the multi-modal framework requires eye-paired datasets of both modalities, which is impractical for clinical use. 
To address this problem, we propose a novel fundus-enhanced disease-aware distillation model (\textbf{FDDM}), for retinal disease classification from OCT images. 
Our framework enhances the OCT model during training by utilizing unpaired fundus images and does not require the use of fundus images during testing, which greatly improves the practicality and efficiency of our method for clinical use.
Specifically, we propose a novel class prototype matching to distill disease-related information from the fundus model to the OCT model and a novel class similarity alignment to enforce consistency between disease distribution of both modalities.  Experimental results show that our proposed approach outperforms single-modal, multi-modal, and state-of-the-art distillation methods for retinal disease classification. Code is available at \href{https://github.com/xmed-lab/FDDM}{https://github.com/xmed-lab/FDDM}.

\end{abstract}


\section{Introduction}
Retinal diseases are one of the most common eye disorders, which can lead to vision impairment and blindness if left untreated. 
Computer-aided diagnosis has been increasingly used as a tool to detect ophthalmic diseases at the earliest possible time and to ensure rapid treatment. 
Optical Coherence Tomography (OCT) \cite{huang1991optical} is an innovative imaging technique with the ability to capture micrometer-resolution images of retina layers, which provides a deeper view compared to alternative methods, such as fundus photographs~\cite{muller2019ophthalmic}, thereby allowing diseases to be detected earlier and more accurately. 
Because of this, OCT imaging has become the primary diagnostic test for many diseases, such as age-related macular degeneration, central serous chorioretinopathy, and retinal vascular occlusion~\cite{ehlers2019retina}.

\begin{figure}[t]
	\centering
	\includegraphics[width=0.9\textwidth]{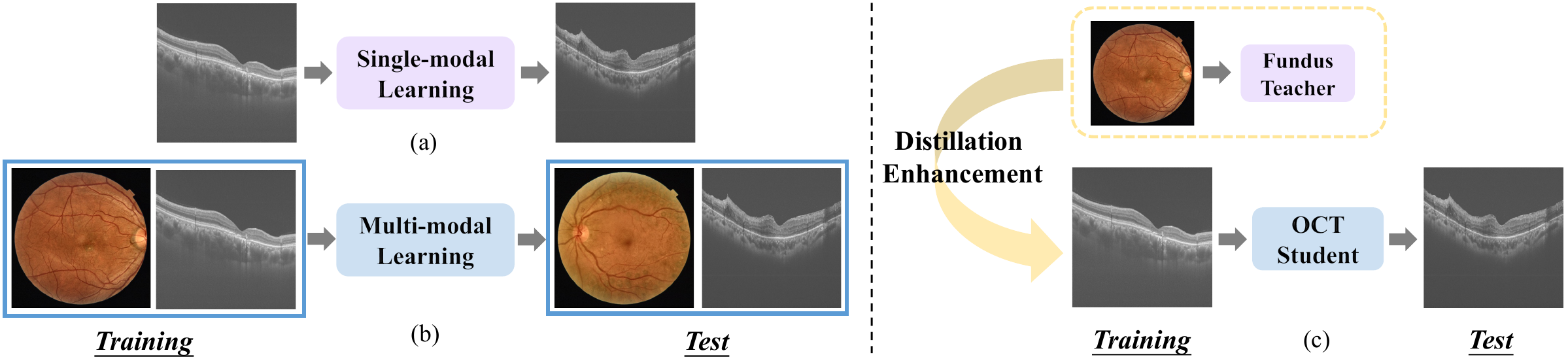}
	\caption{Comparison between (a) single-modal, (b) multi-modal learning, and (c) our proposed distillation enhancement method. Under our setting, images from an additional modalitiy are only used for model training and are not required for inference.}
	\label{fig:setting}
\end{figure}

Traditional methods manually design OCT features and adopt machine learning classifiers for prediction \cite{liu2011automated,srinivasan2014fully,lemaitre2016classification}. 
In recent years, deep learning methods have achieved outstanding performance on various medical imaging analysis tasks and have also been successfully applied to retinal disease classification with OCT images \cite{lee2017deep,karri2017transfer,kermany2018identifying}.
However, diagnosing disease with a single OCT modality, as shown in Fig.~\ref{fig:setting} (a), is still challenging since OCT scans are inadequate compared with fundus photos due to their more expensive cost in data collection.
Some methods attempt to use extra layer-related knowledge from the segmentation task to improve prediction despite limited OCT data \cite{li2019deep,huang2019automatic,fang2019attention,liu2021one}, but this leads to increased training costs since an additional segmentation model is required. 


Recent works have attempted to include additional modalities for classification through multi-modal learning shown in Fig.~\ref{fig:setting} (b),
where fundus and OCT images are jointly used to detect various retinal diseases and achieve promising results~\cite{yoo2019possibility,wang2019two,wang2022learning,ou2021m,he2021multi,li2021multi,li2022multimodal}.
Wang \textit{et al.} \cite{wang2019two,wang2022learning} used a two-stream structure to extract fundus and OCT features, which are then concatenated for prediction. He \textit{et al.} \cite{he2021multi} designed modality-specific attention networks to tackle differences in modal characteristics.
Nevertheless, there are still limitations in these existing approaches.
Firstly, existing multi-modal learning approaches require strictly paired images from both modalities for training and testing. This necessitates the collection of multi-modal images for the same patients, which can be laborious, costly, and not easily achievable in real-world clinical practice.
Secondly, previous works mostly focused on a limited set of diseases, such as age-related macular degeneration (AMD), diabetic retinopathy, and glaucoma, which cannot reflect the complexity and diversity of real-world clinical settings.  

To this end, we propose \textbf{F}undus-enhanced \textbf{D}isease-aware \textbf{D}istillation \textbf{M}odel (\textbf{FDDM}) for retinal disease classification from OCT images, as shown in Fig.~\ref{fig:setting} (c). 
FDDM is motivated by the observation that fundus images and OCT images provide complementary information for disease classification. For instance, in the case of AMD detection, fundus images can provide information on the number and area of drusen or atrophic lesions of AMD, while OCT can reveal the aggressiveness of subretinal and intraretinal fluid lesions~\cite{yoo2019possibility}. Utilizing this complementary information from both modalities can enhance AMD detection accuracy. 

Our main goal is to extract disease-related information from a fundus teacher model and transfer it to an OCT student model, all without relying on paired training data. To achieve this, we propose a class prototype matching method to align the general disease characteristics between the two modalities while also eliminating the adverse effects of a single unreliable fundus instance.
Moreover, we introduce a novel class similarity alignment method to encourage the student to learn similar inter-class relationships with the teacher, thereby obtaining additional label co-occurrence information. Unlike existing works, our method is capable of extracting valuable knowledge from any accessible fundus dataset without additional costs or requirements. Moreover, our approach only needs one modality during the inference process, which can help \textit{greatly reduce the prerequisites for clinical application}.

To summarize, our main contributions include 1) We propose a novel fundus-enhanced disease-aware distillation model for retinal disease classification via class prototype matching and class similarity alignment; 2) Our proposed method offers flexible knowledge transfer from any publicly available fundus dataset, which can significantly reduce the cost of collecting expensive multi-modal data. This makes our approach more accessible and cost-effective for retinal disease diagnosis; 3) We validated our proposed method using a clinical dataset and other publicly available datasets. The results demonstrate superior performance when compared to state-of-the-art alternatives, confirming the effectiveness of our approach for retinal disease classification.

\section{Methodology}
\label{sec:methodology}
\begin{figure*}[t]
	\centering
	\includegraphics[width=0.95\textwidth]{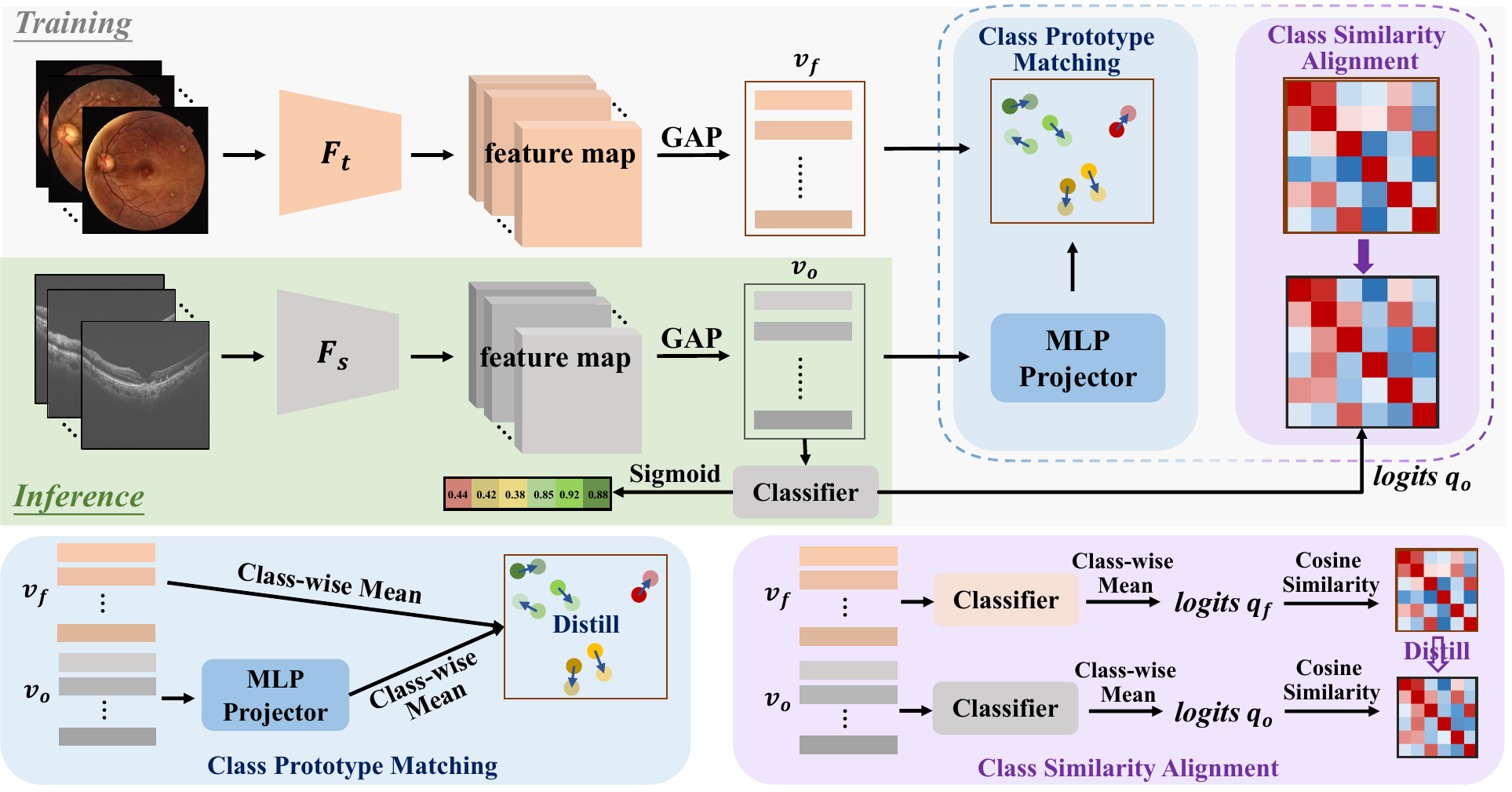}
	\caption{Overview of our proposed FDDM. Our method is based on class prototype matching, which distills disease-specific features, and class similarity alignment, which distills inter-class relationships.
        }
	\label{fig:method}
\end{figure*}

Our approach is based on two ideas: class prototype matching, which distills generalized disease-specific knowledge unaffected by individual sample noise, and class similarity alignment, which transfers additional label co-occurrence information from the teacher to the student. Details of both components are discussed in the sections below. An overview of our framework is shown in Fig.~\ref{fig:method}.


We denote the fundus dataset as $D_f = \{x_{f,i}, y_{f,i}\}_{i=1}^N$, and the OCT dataset as  $D_o = \{x_{o,j}, y_{o,j}\}_{j=1}^M$. To utilize knowledge from the fundus modality during training, we build a teacher model, denoted $F_t$, trained on $D_f$. Similarly, an OCT model $F_s$ is built to learn from OCT images $D_o$ using the same backbone architecture as the fundus model. We use binary cross-entropy loss as the classification loss $\mathcal{L}_{CLS}$ for optimization, to allow the same input to be associated with multiple classes. During inference time, only OCT data is fed into the OCT model to compute the probabilities $\boldsymbol{p}=\{p^c\}_{c=1}^C$ for each disease, $c$. 

\subsection{Class Prototype Matching}
\label{sec:distill}

To distill features from the teacher model into the student model, we aim to ensure that features belonging to the same class are similar. However, we note that individual sample features can be noisy since they contain variations specific to the sample instance instead of the class. In order to reduce noise and ensure disease-specific features are learnt, 
we compress features of each class into a class prototype vector to represent the general characteristics of the disease. During the training per batch, the class prototype vector is the average of all the feature vectors belonging to each category, which is formulated as:
\begin{equation}
    \boldsymbol{e}^c_f=\frac{\sum_{i=1}^B{v_{f,i}*y_{f,i}^c}}{\sum_{i=1}^B{y_{f,i}^c}}, \ \boldsymbol{e}^c_o=\frac{\sum_{j=1}^B{{\mathbf{P}}\left(v_{o,j}\right)*y_{o,j}^c}}{\sum_{j=1}^B{y_{o,j}^c}} \:,
\end{equation}
where $\boldsymbol{e}^c_f$ and $\boldsymbol{e}^c_o$ denote the prototype vector for class $c$ of the fundus and OCT modality respectively, $v_{f,i}$ $\left( v_{o,j}\right)$ represents the feature vector of the input image, and $y_{f,i}^c$ $\left(y_{o,j}^c\right)$ is a binary number which indicates whether the instance belongs to class $c$ or not. $\mathbf{P}$ demotes an MLP projector that projects OCT features into the same space as fundus features.

In the class prototype matching stage, 
we apply softmax loss to the prototype vectors of fundus modality to formulate soft targets $\mathcal{E}_f^c=\sigma(\boldsymbol{e}_f^c/\tau)$, where $\tau$ is the temperature scale that controls the strength to soften the distribution. 
Student class prototypes $\mathcal{E}_o^c$ are obtained in the same way. KL divergence is then used to encourage OCT student to learn matched class prototypes with fundus teacher:
\begin{equation}
\mathcal{L}_{CPM}=\sum_{c=1}^C{\mathcal{E}_f^c\log{\left(\frac{\mathcal{E}_f^c}{\mathcal{E}_o^c}\right)} \:,}
\end{equation}
By doing so, the OCT model is able to use the global information from fundus modality for additional supervision. Overall, our approach adopts class prototypes from fundus modality instead of noisy features from individual samples, which provides more specific knowledge for OCT student model.

\subsection{Class Similarity Alignment}
We also note that for multi-label classification tasks, relationships among different classes also contain important information, especially since label co-occurrence is common for eye diseases. Based on this observation, we additionally propose a class similarity alignment scheme to distill knowledge concerning inter-class relationships from fundus model to OCT model.

First, we estimate the disease distribution by averaging the obtained logits of fundus and OCT model in a class-wise manner to get $\boldsymbol{q}_f=\{q_f^c\}_{c=1}^C, \boldsymbol{q}_o=\{q_o^c\}_{c=1}^C$.
Then, to transfer information on inter-class relationships, we enforce cosine similarity matrices of the averaged logits to be consistent between teacher and student model. The similarity matrix for teacher model is calculated as $\mathcal{Q}_f^c=\sigma(sim(q_f^c, \boldsymbol{q}_f)/\tau)$ and is obtained similarly for student model, $\mathcal{Q}_o^c$. KL divergence loss 
is used to encourage alignment between the two similarity matrices:
\begin{equation}   \mathcal{L}_{CSA}=\sum_{c=1}^C{\mathcal{Q}_f^c\log{\left(\frac{\mathcal{Q}_f^c}{\mathcal{Q}_o^c}\right)} \:,}
\end{equation}


In this way, disease distribution knowledge is distilled from fundus teacher model, forcing OCT student model to learn additional knowledge concerning inter-class relationships, which is highly important in multi-label scenarios.

\subsection{Overall framework}
The overall loss is the combination of classification loss and distillation enhancement loss:
\begin{equation}
    \mathcal{L}_{OCT}=\mathcal{L}_{CLS} + \alpha\mathcal{L}_{CPM} + \beta\mathcal{L}_{CSA} \:,
\end{equation}
where $\alpha$ and $\beta$ are loss weights that control the contribution of each distillation loss. 
Admittedly, knowledge distillation strategies in computer vision \cite{romero2014fitnets,hinton2015distilling,park2019relational,shu2021channel,zhao2022decoupled,chen2022knowledge} can be applied to share multi-modal information as well. 
Unlike classical distillation methods, our two novel distillation losses allow knowledge about \textit{disease-specific features} and \textit{inter-class relationships} to be transferred, thereby allowing knowledge distillation to be conducted with unpaired data. 


\section{Experiments}
\label{sec:experiments}
\subsection{Experimental Setup}
\subsubsection{Dataset.}

To evaluate the effectiveness of our approach, we collect a new dataset TOPCON-MM with paired fundus and OCT images from 369 eyes of 203 patients in Guangdong Provincial Hospital of Integrated Traditional Chinese and Western Medicine using a Topcon Triton swept-source OCT featuring multimodal fundus imaging. 
For fundus images, they are acquired at a resolution of 2576$\times$1934. For OCT scans, the resolution ranges from 320 $\times$ 992 to 1024 $\times$ 992. 
Specifically, multiple fundus and OCT images are obtained for each eye, and each image may reveal multiple diseases with consistent labels specific to that eye. All cases were examined by two ophthalmologists independently to determine the diagnosis label.
If the diagnosis from two ophthalmologists disagreed with each other, a third senior ophthalmologist with more than 15 years of experience was consulted to determine the final diagnosis.
As shown in Table~\ref{tab:data-statistics}, there are eleven classes, including  normal, dry age-related macular degeneration (dAMD), wet age-related macular degeneration (wAMD), diabetic retinopathy (DR), central serous chorioretinopathy (CSC), pigment epithelial detachment (PED), macular epiretinal membrane (MEM), fluid (FLD), exudation (EXU), choroid neovascularization (CNV) and retinal vascular occlusion (RVO).








\begin{table}[t]
    \caption{Statistics of our collected TOPCON-MM dataset.}
    \label{tab:data-statistics}
    \centering
    \begin{adjustbox}{width=0.76\textwidth}
        \begin{tabular}{|c|c|c|c|c|c|c|c|c|c|c|c|c|}
        \hline
        \textbf{Category} & Normal & dAMD & wAMD & DR & CSC & PED & MEM & FLD & EXU & CNV & RVO & Total \\ 
        \hline
        \textbf{Eyes} & 153 & 52 & 30 & 72 & 15 & 23 & 38 & 93 & 90 & 14 & 10 & 369 \\
        \hline
        \textbf{Fundus Images} & 299 & 178 & 171 & 502 & 95 & 134 & 200 & 638 & 576 & 143 & 34 & 1520 \\
        \hline
        \textbf{OCT Images} & 278 & 160 & 145 & 502 & 95 & 133 & 196 & 613 & 573 & 138 & 34 & 1435 \\
        \hline
        \end{tabular}
    \end{adjustbox}
\end{table}

\subsubsection{Implementation Details.}

Following prior work~\cite{wang2019two,wang2022learning}, we use contrast-limited adaptive histogram equalization for fundus images and median filter for OCT images as data preprocessing. We adopt data augmentation including random crop, flip, rotation, and changes in contrast, saturation, and brightness. 
All the images are resized to 448$\times$448 before feeding into the network.
For a fair comparison, we apply identical data processing steps, data augmentation operations, model backbones and running epochs in all the experiments.
We use SGD to optimize parameters with a learning rate of 1e-3, a momentum of 0.9, and a weight decay of 1e-4. The batch size is set to 8. For weight parameters, $\tau$ is set to 4, $\alpha$ is set to 2 and $\beta$ is set to 1. All the models are implemented on an NVIDIA RTX 3090 GPU. We split the dataset into training and test subsets according to the patient's identity and maintained a training-to-test set ratio of approximately 8:2. To ensure the robustness of the model, the result was reported by five-fold cross-validation.


\subsubsection{Evaluation Metrics.} 
We follow previous work \cite{li2021multi} to evaluate image-level performance. As each eye in our dataset was scanned multiple times, we use the ensemble results from all the images of the same eye to determine the final prediction. More specifically, if any image indicates an abnormality, the eye is predicted to have the disease.

\subsection{Compare with State-of-the-Arts}


\begin{table*}[t]
    \caption{Results on our collected TOPCON-MM dataset.
    ``Training" and ``Inference" indicate which modalities are required for both phases. ``Paired" indicates whether paired fundus-OCT images are required in training. All the experiments use ResNet50 as the backbone. For multi-modal methods, two ResNet50 without shared weights are applied separately for each modality. ``Late Fusion" refers to the direct ensemble of the results from models trained with two single modalities. $\dagger$: we implement multi-modal methods in retinal disease classification. $\star$: we run KD methods in computer vision. 
    }
    \label{tab:main-results}
    \centering
    \begin{adjustbox}{width=0.9\textwidth}
        \begin{tabular}{l|ccc|ccccc}
        \hline
        \textbf{Method} & \textbf{Training}& \textbf{Inference} & \textbf{Paired} & \textbf{MAP} & \textbf{Sensitivity} & \textbf{Specificity} & \textbf{F1 Score} & \textbf{AUC} \\
        \hline
        \multicolumn{9}{c}{\textit{Single-Modal Methods}} \\
        \hline
        ResNet50 & Fundus & Fundus & - & 50.56\scriptsize{$\pm$3.05} & 43.68\scriptsize{$\pm$6.58} & 92.24\scriptsize{$\pm$0.77} & 54.95\scriptsize{$\pm$7.09} & 79.97\scriptsize{$\pm$1.63} \\
        ResNet50 & OCT & OCT & - & 66.44\scriptsize{$\pm$3.81} & 53.14\scriptsize{$\pm$6.60} & 95.28\scriptsize{$\pm$0.85} & 64.16\scriptsize{$\pm$6.24} & 87.73\scriptsize{$\pm$1.44} \\
        \hline 
        \multicolumn{9}{c}{\textit{Multi-Modal Methods}} \\
        \hline
        Late Fusion & Both & Both & \ding{51} & 63.83\scriptsize{$\pm$1.34} & 54.45\scriptsize{$\pm$2.72} & 94.29\scriptsize{$\pm$0.74} & 64.93\scriptsize{$\pm$3.00} & 86.92\scriptsize{$\pm$1.48}  \\
        Two-Stream CNN$\dagger$~\cite{wang2019two} & Both & Both & \ding{51} & 58.75\scriptsize{$\pm$2.71} & 53.47\scriptsize{$\pm$3.82} & 92.97\scriptsize{$\pm$0.91} & 61.82\scriptsize{$\pm$4.02} & 84.79\scriptsize{$\pm$2.77} \\
        MSAN$\dagger$~\cite{he2021multi} & Both & Both & \ding{51} & 59.49\scriptsize{$\pm$3.43} & 56.44\scriptsize{$\pm$3.13} & 93.37\scriptsize{$\pm$0.59} & 63.95\scriptsize{$\pm$3.77} & 84.51\scriptsize{$\pm$1.91} \\ 
        FitNet$\star$~\cite{romero2014fitnets} & Both & OCT & \ding{51} & 63.41\scriptsize{$\pm$3.45} & 54.44\scriptsize{$\pm$4.04} & 94.87\scriptsize{$\pm$0.58} & 65.00\scriptsize{$\pm$4.32} & 87.17\scriptsize{$\pm$1.85} \\
        KD$\star$~\cite{hinton2015distilling} & Both & OCT & \ding{51} & 63.69\scriptsize{$\pm$2.04} & 51.70\scriptsize{$\pm$3.10} & 95.75\scriptsize{$\pm$0.62} & 63.56\scriptsize{$\pm$2.32} & 87.90\scriptsize{$\pm$1.03} \\
        RKD$\star$~\cite{park2019relational} & Both & OCT & \ding{51} & 63.59\scriptsize{$\pm$3.04} & 53.42\scriptsize{$\pm$1.71} & 94.42\scriptsize{$\pm$1.81} & 63.70\scriptsize{$\pm$0.72} & 87.36\scriptsize{$\pm$2.08} \\
        DKD$\star$~\cite{zhao2022decoupled} & Both & OCT & \ding{51} & 64.40\scriptsize{$\pm$2.09} & 53.83\scriptsize{$\pm$5.23} & 95.24\scriptsize{$\pm$0.11} & 64.00\scriptsize{$\pm$4.44} & 87.52\scriptsize{$\pm$0.58} \\
        SimKD$\star$~\cite{chen2022knowledge} & Both & OCT & \ding{51} & 65.10\scriptsize{$\pm$2.63} & 53.13\scriptsize{$\pm$5.49} & 95.09\scriptsize{$\pm$0.92} &  63.19\scriptsize{$\pm$6.53} & 87.97\scriptsize{$\pm$1.32} \\
        Ours & Both & OCT & \ding{55} & \textbf{69.06}\scriptsize{$\pm$3.39} & \textbf{57.15}\scriptsize{$\pm$5.93} & \textbf{95.93}\scriptsize{$\pm$0.57} & \textbf{69.17}\scriptsize{$\pm$6.07} & \textbf{89.06}\scriptsize{$\pm$0.97} \\
        \hline
        \end{tabular}
    \end{adjustbox}
\end{table*}
To prove the effectiveness of our proposed method, we compare our approach with single-modal, multi-modal, and knowledge distillation methods. 
From Table~\ref{tab:main-results},  it is apparent that the model trained with OCT alone performs better than the fundus models.
It is noteworthy that current multi-modality methods~\cite{wang2019two,he2021multi} and knowledge distillation methods~\cite{romero2014fitnets,hinton2015distilling,park2019relational,zhao2022decoupled,chen2022knowledge} do not yield improved results on our dataset. 
Table~\ref{tab:main-results} also demonstrates that compared with the single-modal OCT baseline, our method improves MAP from 66.44\% to 69.06\%, F1 from 64.16\% to 69.17\%.
\textit{This shows that it is still possible to learn valuable information from the fundus modality to assist the OCT model, despite being a weaker modality.} It can be observed that our approach outperforms the state-of-the-art multi-modal retinal image classification method \cite{he2021multi} by 9.57\% in MAP (69.06\% \vs 59.49\%). Notably, our method excels the best-performing knowledge distillation method \cite{chen2022knowledge} by 3.96\% in MAP (69.06\% \vs 65.10\%).
We also note that the alternative methods are limited to training with eye-paired fundus and OCT images only, whilst our approach does not face such restrictions.

To further demonstrate the efficiency of our proposed distillation enhancement approach, we validate our method on a publicly available multi-modal dataset with fundus and OCT images, MMC-AMD \cite{wang2022learning}. MMC-AMD dataset contains four classes: normal, dry AMD, PCV, and wet AMD. We reproduce single-modal ResNet, Two-Stream CNN \cite{wang2019two}, and KD methods\cite{romero2014fitnets,hinton2015distilling} as baselines and show results in Fig.~\ref{fig:other-dataset} (a). It can be seen that our method improves MAP to 92.29\%, largely surpassing existing methods. 



\subsection{Results Trained with Other Fundus Datasets}

Since we implement distillation in a disease-aware manner, multi-modal fundus and OCT training data do not need to be paired. Theoretically, any publicly available fundus dataset could be applied as long as it shares a label space that overlaps with our OCT data. To verify this hypothesis, we separately reproduce our methods with fundus images from two datasets, MMC-AMD \cite{wang2022learning} and RFMiD \cite{pachade2021retinal}. 
To ensure label overlap, we only select fundus and OCT images from common classes for training and validation, namely, 3 classes for MMC-AMD and 6 classes for RFMiD.
The results are reported in Fig.~\ref{fig:other-dataset} (b). Compared with single-modal OCT model, our distillation enhancement can achieve an increase of 4.26\% (84.06\% \vs 79.80\%) and 2.21\% (75.47\% \vs 73.26\%) in MAP. Our results proove that \textit{our method has the flexibility to use any existing fundus dataset to enhance OCT classification}.

\pgfplotstableread[row sep=\\,col sep=&]{
 	Metric & MAP \\
	{Fundus CNN} & 83.90 \\
 {OCT CNN} & 87.98 \\
 {Two-Stream} & 86.91 \\
 {FitNet} & 90.72 \\
 {KD} & 90.29 \\
 {Ours} & 92.29 \\
}\mydata

\pgfplotsset{every axis/.append style={
		label style={font=\normalsize},
		tick label style={font=\normalsize} 
}}

\pgfplotstableread[row sep=\\,col sep=&]{
	Model & OCT CNN & Ours  \\
	{MMC-AMD $\rightarrow$ TOPCON-MM} & 79.80 & 84.06 \\
        {RFMiD $\rightarrow$ TOPCON-MM} & 73.26 & 75.47 \\
}\mydatatransfer

\pgfplotsset{every axis/.append style={
		label style={font=\normalsize},
		tick label style={font=\normalsize} 
}}

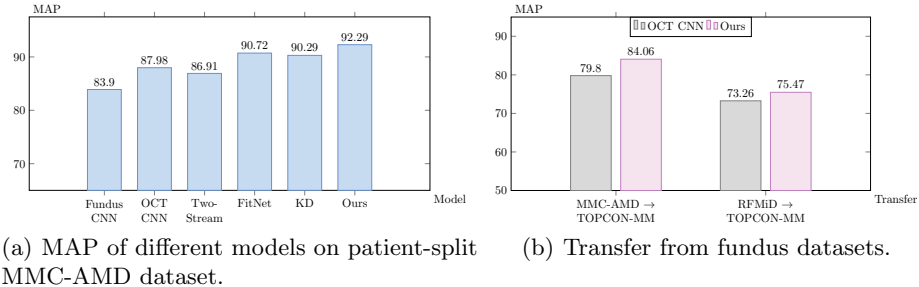
\begin{figure}[t]
    \subfigure[MAP of different models on patient-split MMC-AMD dataset.]{
	\centering
        \begin{tikzpicture}[scale=0.45]
		\tikzstyle{every node}=[font=\normalsize]
		\begin{axis}[
			ybar=8pt,
			bar width=1cm,
			enlarge x limits=0.3,
			width= 1.1 \textwidth,
			height= 0.55 \textwidth,
			symbolic x coords={{Fundus CNN}, {OCT CNN}, {Two-Stream}, {FitNet}, {KD}, {Ours}},
   x tick label style = {text width = 1.5cm, align=center},
			xtick=data,
                xlabel=Model,
                xlabel style={at={(ticklabel* cs:1)},anchor=north west},
                ylabel=MAP,
                 ylabel style={at={(ticklabel* cs:1)},anchor=south west,rotate=-90},
			nodes near coords,
			every node near coord/.append style={font=\normalsize},
			nodes near coords align={vertical},
			ymin=65,ymax=97.5,
			]
			\addplot 
			[draw=cyan!40!blue!60!white,fill=cyan!60!blue!20!white] 
			table[x=Metric,y=MAP]{\mydata};
		\end{axis}
	\end{tikzpicture}
    }
    \subfigure[Transfer from fundus datasets.]{
	\centering
	\begin{tikzpicture}[scale=0.45]
		\tikzstyle{every node}=[font=\normalsize]
		\begin{axis}[
			ybar=8pt,
			bar width=1.2cm,
			enlarge x limits=0.7,
			width= 1.0 \textwidth,
			height= 0.55 \textwidth,
			legend style={at={(0.5,1)},
				anchor=north,legend columns=-1},
			symbolic x coords=
			{{MMC-AMD $\rightarrow$ TOPCON-MM}, {RFMiD $\rightarrow$ TOPCON-MM}},
      x tick label style = {text width = 4cm, align=center},
			xtick=data,
                xlabel=Transfer,
                xlabel style={at={(ticklabel* cs:1)},anchor=north west},
                ylabel=MAP,
                 ylabel style={at={(ticklabel* cs:1)},anchor=south west,rotate=-90},
			nodes near coords,
			every node near coord/.append style={font=\normalsize},
			nodes near coords align={vertical},
			ymin=50,ymax=95,
			]
			\addplot 
			[draw=gray,fill=gray!30!white] 
			table[x=Model,y=OCT CNN]{\mydatatransfer};
                \addplot 
			[draw=violet!50!white,fill=violet!40!pink!20!white] 
			table[x=Model,y=Ours]{\mydatatransfer};
			\legend{{OCT CNN}, {Ours}}
		\end{axis}
	\end{tikzpicture}
    }
	\caption{Results on other publicly available datasets.}
 \label{fig:other-dataset}
\end{figure}

\subsection{Ablation Studies}

Table~\ref{tab:ablation} shows the ablation study of our method. 
To provide additional insight, we also show the results on majority classes, which contain over 10\% images of the dataset, and minority classes with less than 10\%. 
It can be seen that individually using CPM and CSA can improve the overall result by $1.32\%$ and $1.06\%$ in MAP, respectively. Removing either of the components degrades the performance. 
Results also show that CPM improves classification performance in majority classes by distilling disease-specific knowledge, while CSA benefits minority classes by attending to inter-disease relationships. By simultaneously adopting CPM and CSA, the overall score of all the classes is improved. 






\begin{table}[t]
    \caption{Ablation study of our method. ``Majority" and ``Minority" refers to the average score of classes that represent more than 10\% or less than 10\% of the total number of images, respectively. ``Overall" indicates overall performance on all the classes.}
    \label{tab:ablation}
    \centering
    \begin{adjustbox}{width=0.7\textwidth}
        \begin{tabular}{cc|ccc|c|c}
        \hline
        \multicolumn{2}{c|}{\textbf{Method}} & \multicolumn{3}{c|}{\textbf{MAP}} & \multirow{2}{*}{\textbf{F1 Score}} & \multirow{2}{*}{\textbf{AUC}} \\
        \cline{3-5}
        \textbf{CPM} & \textbf{CSA} & Overall & Majority & Minority & & \\
        \hline
        \ding{55} & \ding{55} & 66.44\scriptsize{$\pm$3.81} & 71.12\scriptsize{$\pm$4.04} & 58.26\scriptsize{$\pm$7.42} & 64.16\scriptsize{$\pm$6.24} & 87.73\scriptsize{$\pm$1.44} \\
        \ding{55} & \ding{51} & 67.50\scriptsize{$\pm$3.00} & 70.60\scriptsize{$\pm$4.91} & \textbf{62.08}\scriptsize{$\pm$8.72} & 65.40\scriptsize{$\pm$5.21} & 88.09\scriptsize{$\pm$1.28} \\
        \ding{51} & \ding{55} & 67.76\scriptsize{$\pm$2.34} & 72.26\scriptsize{$\pm$3.20} & 59.90\scriptsize{$\pm$8.32} & 65.58\scriptsize{$\pm$2.58} & 88.73\scriptsize{$\pm$1.32} \\\
        \ding{51} & \ding{51} & \textbf{69.06}\scriptsize{$\pm$3.39} & \textbf{73.34}\scriptsize{$\pm$3.48} & 61.47\scriptsize{$\pm$8.17} & \textbf{69.17}\scriptsize{$\pm$6.07} & \textbf{89.06}\scriptsize{$\pm$0.97} \\
        \hline
        \end{tabular}
    \end{adjustbox}
\end{table}



\section{Conclusion}
Our work proposes a novel fundus-enhanced disease-aware distillation module, FDDM, for retinal disease classification. The module incorporates class prototype matching to distill global disease information from the fundus teacher to the OCT student, while also utilizing class similarity alignment to ensure the consistency of disease relationships between both modalities. Our approach deviates from the existing models that rely on paired instances for multi-modal training and inference, making it possible to extract knowledge from any available fundus data and render predictions with only OCT modality. As a result, our approach significantly reduces the prerequisites for clinical applications. Our extensive experiments demonstrate that our method outperforms existing baselines by a considerable margin.

\vspace{6pt}
\noindent
\textbf{Acknowledgement.} This work is supported by grants from Foshan HKUST Projects under Grants FSUST21-HKUST10E and FSUST21-HKUST11E, as well as by the Hong Kong Innovation and Technology Fund under Projects PRP/041/22FX and ITS/030/21.





%
%

\end{document}